\documentclass{svmult}

\usepackage{epsfig}
\usepackage{graphicx}
\usepackage{rotate}
\usepackage{epsf}
\usepackage{natbib}
\usepackage{makeidx}         
\usepackage{graphicx}        
\usepackage{multicol}        
\usepackage[bottom]{footmisc}

\makeindex             

\begin{document}
\def\aj{AJ}
\def\araa{ARA\&A}
\def\apj{ApJ}
\def\apjl{ApJ}
\def\apjs{ApJS}
\def\ao{Appl.~Opt.}
\def\apss{Ap\&SS}
\def\aap{A\&A}
\def\aapr{A\&A~Rev.}
\def\aaps{A\&AS}
\def\azh{AZh}
\def\baas{BAAS}
\def\jrasc{JRASC}
\def\memras{MmRAS}
\def\mnras{MNRAS}
\def\pra{Phys.~Rev.~A}
\def\prb{Phys.~Rev.~B}
\def\prc{Phys.~Rev.~C}
\def\prd{Phys.~Rev.~D}
\def\pre{Phys.~Rev.~E}
\def\prl{Phys.~Rev.~Lett.}
\def\pasp{PASP}
\def\pasj{PASJ}
\def\qjras{QJRAS}
\def\skytel{S\&T}
\def\solphys{Sol.~Phys.}
\def\sovast{Soviet~Ast.}
\def\ssr{Space~Sci.~Rev.}
\def\zap{ZAp}
\def\nat{Nature}
\def\iaucirc{IAU~Circ.}
\def\aplett{Astrophys.~Lett.}
\def\apspr{Astrophys.~Space~Phys.~Res.}
\def\bain{Bull.~Astron.~Inst.~Netherlands}
\def\fcp{Fund.~Cosmic~Phys.}
\def\gca{Geochim.~Cosmochim.~Acta}
\def\grl{Geophys.~Res.~Lett.}
\def\jcp{J.~Chem.~Phys.}
\def\jgr{J.~Geophys.~Res.}
\def\jqsrt{J.~Quant.~Spec.~Radiat.~Transf.}
\def\memsai{Mem.~Soc.~Astron.~Italiana}
\def\nphysa{Nucl.~Phys.~A}
\def\physrep{Phys.~Rep.}
\def\physscr{Phys.~Scr}
\def\planss{Planet.~Space~Sci.}
\def\procspie{Proc.~SPIE}


\newcommand{\Lsolar}{L$_{\odot}$}
\newcommand{\farcmin}{\hbox{$.\mkern-4mu^\prime$}}
\newcommand{\farcsec}{\hbox{$.\!\!^{\prime\prime}$}}
\newcommand{\kms}{\rm km\,s^{-1}}
\newcommand{\cc}{\rm cm^{-3}}
\newcommand{\Alfven}{$\rm Alfv\acute{e}n$}
\newcommand{\Vap}{V^\mathrm{P}_\mathrm{A}}
\newcommand{\Vat}{V^\mathrm{T}_\mathrm{A}}
\newcommand{\Msolar}{$\,{\cal M}_{\odot}$}
\newcommand{\Mstar}{$\,{\cal M}_{\star}$}
\newcommand{\Mdot}{\,\dot{\cal M}}
\newcommand{\beq}{\begin{equation}}
\newcommand{\eeq}{\end{equation}}
\newcommand{\ben}{\begin{enumerate}}
\newcommand{\een}{\end{enumerate}}
\newcommand{\bit}{\begin{itemize}}
\newcommand{\eit}{\end{itemize}}
\newcommand{\bec}{\begin{center}}
\newcommand{\enc}{\end{center}}
\newcommand{\barr}{\begin{array}}
\newcommand{\earr}{\end{array}}
\newcommand{\DD}{\frac}
\newcommand{\ber}{\begin{array}}
\newcommand{\eer}{\end{array}}
\newcommand{\LLRA}{\Longleftrightarrow}
\newcommand{\LRA}{\Leftrightarrow}
\newcommand{\llra}{\longleftrightarrow}
\newcommand{\lra}{\leftrightarrow}
\newcommand{\DRA}{\Downarrow}
\newcommand{\diver}{\mbox{\,div}}
\newcommand{\grad}{\mbox{\,grad}}
\newcommand{\TAW}{\tiny{\rm TAW}}
\newcommand{\mm }{\mathrm}
\newcommand{\Bp }{B_\mathrm{p}}
\newcommand{\Bpr }{B_\mathrm{r}}
\newcommand{\Bpz }{B_\mathrm{\theta}}
\newcommand{\Bt }{B_\mathrm{T}}
\newcommand{\Vp }{V_\mathrm{p}}
\newcommand{\Vpr }{V_\mathrm{r}}
\newcommand{\Vpz }{V_\mathrm{\theta}}
\newcommand{\Vt }{V_\mathrm{\varphi}}
\newcommand{\Ti }{T_\mathrm{i}}
\newcommand{\Te }{T_\mathrm{e}}
\newcommand{\rtr }{r_\mathrm{tr}}
\newcommand{\rbl }{r_\mathrm{BL}}
\newcommand{\rtrun }{r_\mathrm{trun}}
\newcommand{\thet }{\theta}
\newcommand{\thetd }{\theta_\mathrm{d}}
\newcommand{\thd }{\theta_d}
\newcommand{\thw }{\theta_W}
\newcommand{\eps}{\epsilon}
\newcommand{\veps}{\varepsilon}
\newcommand{\vepsdi}{{\cal E}^\mathrm{d}_\mathrm{i}}
\newcommand{\vepsde}{{\cal E}^\mathrm{d}_\mathrm{e}}
\newcommand{\lraS}{\longmapsto}
\newcommand{\Equival}{\Longleftrightarrow}
\newcommand{\cd}{\!\cdot\!}
\newcommand{\Msun}{{\,{\cal M}_{\odot}}}
\title*{
Problem-orientable numerical algorithm for modelling multi-dimensional 
        radiative MHD flows in astrophysics
         {\bf -- the hierarchical solution scenario}
}

\titlerunning{Hierarchical scenario: Radiative MHD solvers}  
\author{Ahmad A. Hujeirat}
\authorrunning{Hujeirat, A.}  
\institute{Applied Mathematics, Universit\"at Heidelberg, 69120 Heidelberg, Germany\\  
\texttt{ahmad.hujeirat@iwr.uni-heidelberg.de}}
%
%
\maketitle

{\underline{\bf Abstract:}  We present a hierarchical approach for
enhancing the robustness of numerical solvers for modelling
  radiative MHD flows in multi-dimensions. 

This approach is based on clustering the entries of the global Jacobian in a
hierarchical manner that enables employing a variety of solution procedures
ranging from a purely explicit time-stepping up to fully implicit schemes.\\
A gradual coupling of the radiative MHD equation with the radiative transfer
equation in higher dimensions is possible.  \\
Using this approach,  it is possible to follow the evolution of strongly time-dependent flows
with low/high accuracies and with efficiency comparable to explicit methods, as well as
 searching quasi-stationary solutions for highly viscous flows. \\
In particular, it is shown that the hierarchical approach is
capable of modelling the formation of jets in active galactic
nuclei and reproduce the corresponding spectral energy distribution  with a reasonable accuracy. \\

   }
\section{Introduction}
Within the last two decades, a tremendous progress has been made
in both computational fluid dynamics (CFD) algorithms and the 
computer hardware technologies. The computing speed and memory
capacity of computers have increased exponentially during this
period. Similarly is in astrophysical fluid dynamics (AFD), which
is   a  rapidly growing research field, and in which modern
numerical methods are extensively used to model the evolution of
rather complicated flows. Unlike CFD, in which implicit methods
are frequently used, the majority of the methods used in AFD are
explicit. Several of them became very popular, e.g., ZEUS \cite{Stone},
 NIRVANA+ \cite{Ziegler}, FLASH \cite{Fryxell}, 
VAC \cite{Toth}, THARM \cite{Gammie}.
The popularity of explicit methods arises from their being easy to
construct, vectorizable, parallelizable and even more efficient as
long as  dynamical evolutions of compressible flows are
concerned. Specifically, for modeling the dynamical evolution of
HD-flows in two and three dimensions explicit methods are highly
superior to-date. For modelling relativistic flows, Koide and
collaborators  \cite{Koide1, Koide2} and \cite{Meier1, Komissarov}
 have developed pioneering general
relativistic MHD solvers. A rather complete review of numerical
approaches for relativistic fluid dynamics is given in \cite{Mart, Font}. A
ZEUS-like scheme for general relativistic MHD has also been
developed and is described in \cite{Villiers}.

These methods, however, are numerically stable as far as the
Courant-Friedrich-Levy number is smaller than unity.  The
corresponding time step size decreases  dramatically with the
incorporation of real astrophysical effects.  Specifically, they
may even stagnate if self-gravity, radiative and chemical effects
are included. Moreover,  explicit methods break down if the flow
is weakly or strongly incompressible, and if the domain of
calculations is subdivided into a strongly stretched mesh.  In an
attempt to enhance their robustness, several alternatives have
been suggested, such as semi-explicit, semi-implicit or even
implicit-explicit methods \cite{Kley, Toth}.
 Nevertheless, their rather limited range of applications
has lead to the fact that most  of the interesting astrophysical
problems remained, indeed, not really solved. A simple example is
the evolution of a steady turbulent accretion disk. It was found
by Balbus \& Hawley (1991) that weak magnetic fields in accretion
disks are amplified, generate turbulence, which in turn
 redistribute the angular momentum in the disk. However, whether this
instability  leads to the long-sought global steady accretion
rate, or is it just a transient phenomenon in which the generation
of turbulence is subsequently suppressed by dynamo action are not
at all clear. Other notable phenomena are the formation and
acceleration of the observed superluminal jets in quasars and in
microquasars, the origin of the quasi-periodic oscillation in low
mass X-ray
binaries or the progenitors of gamma ray burst are still spectacular.

Explicit methods rely on time-extrapolation procedures for
advancing the solution in time. However, in order to provide
physically consistent solutions, it is necessary that these
procedures are numerically stable.  The usual approach for
examining  the stability of numerical methods is to perform the so
called von Neumann analysis \cite[see][for further details]{Hirsch}.
This yields the so called Courant-Friedrich-Levy condition (CFL)
which is known to limit the range of application and severely
affects the robustness of explicit methods. In particular,
equations corresponding to  physical processes occurring on much
 shorter time scales than the hydro-time scale (e.g., radiation,
self-gravitation and chemical reactions) cannot be followed
explicitly. Furthermore, these methods are not suited for
searching solutions that correspond to evolutionary phases
occurring on time scales much longer than the hydro-time scale.
Using high performance  computers to perform a large number of
explicit time steps may lead to accumulation of round-off errors
that can easily distort the propagation of information from the
boundaries and cause divergence of the solution procedure,
 especially if Neumann type conditions are imposed at the boundaries.
\begin{figure}
\begin{center}
{\hspace*{-0.0005cm}
\includegraphics*[width=6.35cm,bb=14 345 567 803,clip]{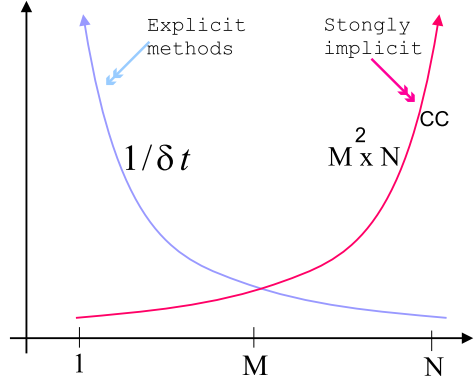}}
\end{center}
{\vspace*{-0.0cm}} \caption [ ] {A schematic description of the
time step size  and the computational costs versus the band width
$M$ of the Jacobian. N is the number of unknowns. Explicit methods
correspond to $M=1$ and large $1/\delta t$. They require minimum
computational costs (CC). Large time steps
 (i.e., small $1/\delta t$) can be achieved using strongly implicit methods.
These methods generally rely on the inversion of matrices with large band
width, hence computationally expensive, and, in most cases,
 are inefficient.
  }
\end{figure}
In contrast to explicit methods, implicit methods are based on
solving a matrix equation of the form $A x=b$,
where ${A}$  is the Jacobian matrix corresponding to  the
system of equations to be solved, $b$ is the
right hand side vector of  known quantities, and $x$ is the
solution vector sought.
 These methods have  two
major drawbacks. First, constructing the matrix A is difficult,
time consuming, and  may considerably influence the robustness of
the method. Second, the inversion procedure  must be stable and
efficient. In general, conservative discretization  of the MHD
equations give rise to sparse matrices, or even to  narrow band
matrices. Therefore, any efficient  matrix inversion procedure
must take the advantage of A being sparse. Inverting A directly by
using  Gaussian elimination requires ${\rm N^{3}}$ algebraic
operations, where N is the number of unknowns. If the flow is
multi-dimensional and a high spatial resolution is required, the
number of operations can be prohibitive even on modern
supercomputers. Krylov Sub-Iterative Methods (KSIMs), on the other
hand, are most suited for sparse matrices and avoid the fill-in
procedure. In the latter case,  A is not directly involved in the
process, but rather its multiplication with a vector. The
convergence rate of KSIMs has been found to depend strongly on the
proper choice of the pre-conditioner. For advection-dominated
flows, incomplete factorization such as ILU, IC and LQ,
approximate factorization, ADI, line Gauss-Seidel are only a small
sub-set of possible sequential pre-conditioners  \cite[see][and the
 references therein]{Saad}. Another powerful way of accelerating relaxation techniques is to use the
multi-grid method as a direct solver or as  a pre-conditioner \cite{Brandt, Trottenberg}.
 For parallel computations, Red-Black ordering in combination with GRMES
and Bi-CGSTAB as well as domain decomposition are among the
popular pre-conditioners \cite[see][for further discussion.]{Dongarra} \\

 Towards studying the jet-disk-BH connection in AGNs and $\mu-$QSOs
a series of multi-dimensional calculations have been performed
\cite[e.g.,][]{Ouyed, Uchida, Meier2, HUJ2, HUJ5, HUJ6}. Specifically, these studies revealed that:
\ben
\item Counter-rotating disks with respect to the BH-spin  generate
      jets that propagate approximately twice as fast as in the co-rotating case.
\item Jets formed are found to be relatively slow, i.e., the corresponding
      $\Gamma-$factors did not reach the desired large values. This was found
      in both cases: when the spins of the disk and the BH are parallel and when they are
      anti-parallel.
      Moreover, disks surrounding Kerr BHs have been verified to produce jets
      that are more powerful than in the  Schwarzschild case. These jets
      are   driven primarily  by strong MFs that are created by the frame dragging
      effect.
\item Large $\Gamma-$factors are obtainable if the {\Alfven} speed due to the PMF
      is equal to or even larger than the local escape velocity 
      \cite[see][and the references therein]{Meier2}.
\item Poliodal magnetic fields may extract rotational energy from the disk plasma, and   from a geometrically thin super-Keplerian layer between the disk and
the overlying corona. The outflowing plasma in this layer is
dissipative, two-temperature, virial-hot, advective and
electron-proton dominated. The innermost part of the disk in this
model is turbulent-free, sub-Keplerian rotating and
advective-dominated. This part  ceases to radiate as a standard
disk, and most of the accretion energy is converted into magnetic
and kinetic energies that go into powering the jet.
 \een

     Nevertheless, jet-structures, their formation, acceleration, their linkage to the accretion
     phenomena and the nature of their plasma are still a matter of debate. Furthermore,
     the flood of observational data makes it even more essential than ever to
     perform  sophisticated numerical calculations to gain a more precise insight of their evolution.

In this paper we focus on the architecture of the global solution procedure
 rather than on local details, such as order of accuracies, physical consistency,
types of advection schemes or fulfilling the solenoidal condition.
Specifically, we discuss strategies for enhancing the robustness
of solvers through constructing various pre-conditionings to
implement  a variety of solution methods in arbitrary dimensions.
Special attention is given to radiative MHD solvers and their
possible coupling with the radiative transfer equation in higher
dimensions.
\section{The governing equations }
\subsection{The 3D axi-symmetric radiative MHD equations   }
Spherical geometry is the most appropriate geometry for capturing
flow configurations   in the vicinity of black holes. Taking into
account the perfect axi-symmetry of black holes, and that their
gravitational pull dominates the forces exerting on the surrounding
flows, we conclude that axi-symmetry is  a reasonable assumption
that may characterize accretion flows in their vicinties.
Moreover, in applying spherical geometry the transformation
$\bar{\theta} = \pi/2 - \theta$ has been used\footnote{This
transformation allows simple analogy with and into cylindrical
coordinates.}. We note that the dynamical time scale near the
event horizon is extremely short,  therefor giving rise to
multi-component flows, such as
electron and ion plasmas.\\
In the following we describe the  set of radiative MHD equations,
and list the scaling variable that may be used for transforming
them into non-dimensional form (see Table 1).

\bit
\item Continuity equation: \beq
   \DD{\partial \rho}{\partial t} +   \nabla \cdot \rho V = 0
\eeq
\item Radial momentum  equation:
    \beq
     \DD{\partial m}{\partial t} +  \nabla \cdot m V
    = \DD{\partial P}{\partial r} + {\rho}\DD{{(\Vpz^2+\Vt}^2)}{r} +\rho \DD{\partial \Phi}{\partial r}
      +  \lambda_\mm{FLD}  \DD{\partial E}{\partial r}  + F^\mm{r}_\mm{L} + Q^\mm{r}_\mm{vis}
\eeq
\item Vertical momentum  equation:
  \beq \DD{\partial n}{\partial t} +   \nabla \cdot n V
    = \DD{\partial P}{\partial \theta} - {\rho {\Vt^2}}\tan{\theta}   + \rho \DD{\partial \Phi}{\partial \theta}
         +  \lambda_\mm{FLD}  \DD{\partial E}{\partial \theta} +  F^\mm{\theta}_\mm{L}
         + Q^{\theta}_\mm{vis}
\eeq
\item Angular momentum  equation:
 \beq \DD{\partial \ell }{\partial t}  + \nabla \cdot \ell V
        =    F^\mm{\varphi}_\mm{L}   + Q^{\varphi}_\mm{vis}
\eeq
\item Internal  equation of the ions:
\beq
   \DD{\partial \vepsdi }{\partial t}   + \nabla \cdot  \vepsdi V
    = -(\gamma -1) \vepsdi \nabla \cdot V
    +   \Phi - \Lambda_\mm{i-e}
    + \nabla \cdot \kappa^\mathrm{cond}_\mathrm{i} \nabla \Ti
\eeq
\item Internal  equation of the electrons:
\beq
   \DD{\partial \vepsde }{\partial t}  + \nabla \cdot  \vepsde V
    = -(\gamma -1) \vepsde \nabla \cdot V
    +  \Lambda_\mm{i-e} - \Lambda_\mm{\rm B }
    -  \Lambda_\mm{\rm C } - \Lambda_\mm{\rm Syn }
    + \nabla \cdot \kappa^\mathrm{cond}_\mathrm{e} \nabla \Te
\eeq
\item
Equation of the zero moment of the  radiation field:
\beq
   \DD{\partial {E} }{\partial t}  + \nabla \cdot  E V
    +  \nabla\cdot[{\lambda_\mm{FLD}}{ \nabla {E}}] - \Lambda_\mm{B} + \Lambda_\mm{C}
   +  \Lambda_\mm{Syn}
\eeq
\item
The induction equation:

\beq
  \DD{\partial B}{\partial t} =  \nabla \times ( V \times B  + \alpha_\mm{dyn} B - \nu_\mm{mag} \nabla \times B). \eeq

\item
Gravitational potential: the Poisson equation:

\beq
  \Delta \psi = 4 \pi G \rho,
 \eeq
where $\psi$ is the gravitational potential and G is the
gravitational constant. \eit In Table (2) we list part of the
variables used and their definitions.
\begin{table*}
\begin{tabular}{ll}
Scaling variables: &   \\\hline
Mass: &   $\tilde{\cal M} = 3\times 10^8M_{\odot}$   \\
Accretion rate: &   $\tilde{\Mdot} = 10^{-1}\Mdot_{Edd}$   \\
Distance: & $\tilde{R} = R_\mm{in} = 3R_\mm{S},$ where $R_\mm{S} = 2G\tilde{\cal M}/c^2$ \\
Temperature: &  $\tilde{\cal T} = 5 \times 10^7 K $  \\
Velocities: &  $\tilde{V} = \tilde{V_\mm{S}} =
       [\gamma {\cal R}_\mm{gas} \tilde{\cal T}/\mu_\mm{i}]^{1/2},\, \mu_\mm{i} = 1.23 $ \\
Ang. Velocity: & $\tilde{\Vt} = {\tilde{V}}_\mm{Kep} = (G\tilde{\cal
M}/\tilde{R})^{1/2}$ \\
Magnetic Fields: & $ \tilde{B} = \tilde{V_\mm{S}}/\sqrt{4 \pi \tilde{\rho}} $  \\
Density: &  $ \tilde{\rho} = \tilde{\Mdot}/(\tilde{H_\mm{d}}
\tilde{R_\mm{out}} \tilde{V_\mm{S}})$ $ = 2.5 \times 10^{-12}
g\,cm^{-3} $ \\\hline
\end{tabular}
\caption{Scaling variables for reformulating the MHD equations in
non-dimensions.}
\end{table*}
Further, the subscripts ``i'' and ``e'' correspond to ion and
electron plasmas,  where $\gamma=5/3$, $\mu_{\rm i}=1.23$ and $
\mu_{\rm e}=1.14 $ are used. $\alpha_\mm{dyn},\, \eta_\mm{mag}$
correspond to the $\alpha-$dynamo and the magnetic diffusivity,
respectively. The radiative diffusion coefficient
$\lambda_\mathrm{FLD}$ is a radiative flux limiter which forces
the radiative flux to adopt the correct form in optically thin and
thick regions, i.e.,
\begin{equation}
{\nabla  \cdot {\lambda_{\rm FLD}}\nabla E =
        \left\{ \begin{array} {c@{\quad \quad}l}
  \nabla \cdot \DD{1}{3\chi}\nabla E  & {\sf if} \hspace*{0.75cm} \tau \gg 1  \\
  \nabla \cdot nE  & {\sf if} \hspace*{0.75cm} \tau \ll 1,
        \end{array} \right. }
\end{equation}
and provides a smooth matching in the transition regions. Here
 $\chi = \rho(\kappa_\mathrm{abs} + \sigma)$ and ${\rm n = \nabla E/|\nabla E|},$
where $\kappa_\mathrm{abs}$ and $\sigma$ are the absorption and
scattering  coefficients. $\Lambda_{B}$, $\Lambda_{\rm i-e}$,
$\Lambda_{\rm C}$, $\Lambda_{\rm syn}$ correspond to
Bremsstrahlung cooling, Coulomb coupling between the ions and
electrons, Compton and synchrotron coolings, respectively \cite{Rybiki}. These
processes read:
\begin{eqnarray}
 \Lambda_{\rm i-e} &=& 5.94\times 10^{-3} n_i n_e c
      k \DD{(T_{\rm i} - T_{\rm e})}{T^{3/2}_{\rm e}} /{\cal N} \nonumber\\
\Lambda_{\rm B} & = &  4ac \kappa_\mathrm{abs} \rho (T^4 - E)/{\cal N},  \\
\Lambda_{\rm C} &=& 4 \sigma n_e c  (\DD{k}{m_e c^2}) (T_e - T_{rad})E/{\cal N}, \nonumber
\end{eqnarray}
where ${\cal N} = [(\gamma-1)/\gamma]
(\tilde{V}^2 \tilde{\Vt}/\tilde{R})$ is a normalization quantity. $n_{\rm e},~n_{\rm i}$ are the electron- and
ion-number densities. E is the density of the radiative energy,
i.e., the zero-moment of the radiative field. The radiative
temperature  is defined as $T_{rad}=E^{1/4}$.
The Lorenz forces acting on charged plasma in the MHD approximation read:
\begin{eqnarray}
F^{r}_\mm{L}  & = &
     \DD{\Bpz}{r} \DD{\partial  \Bpr }{\partial \theta}
    - \DD{1}{r} \DD{\partial  }{\partial r} r (\Bpz^2 + \Bt^2)
    + \DD{1}{2} \DD{\partial }{\partial r}   (\Bpz^2 + \Bt^2 ) \nonumber \\
F^{\theta}_\mm{L}  & = &   {\Bpr} \DD{\partial  }{\partial r} r \Bpz
    - \DD{1}{2 }  \DD{\partial}{\partial \theta} \Bpr^2
- [\DD{1}{\cos{\theta}} \DD{\partial  }{\partial \theta}  {\cos}{\theta}\,\Bt^2
    - \DD{1}{2 } \DD{\partial  }{\partial \theta}  \Bt^2 ]\\
F^{\varphi}_\mm{L}  & = &  \Bp\cdot \nabla \bar{B} =\Bpr \DD{\partial }{\partial r} (r \cos{\theta} \Bt)
   +  \DD{\Bpz}{r} \DD{\partial }{\partial \theta} (r \cos{\theta} \Bt).    \nonumber
\end{eqnarray}
The turbulent-diffusive terms read:
\begin{eqnarray}
Q^{r}_\mm{vis}  & = &
\DD{1}{r^2} \DD{\partial}{\partial r} (r^2 T_\mm{rr})
 + \DD{1}{r \cos{\theta} }\DD{\partial  }{\partial \theta} (\cos{\theta} T_\mm{r \theta})
 + \DD{T_\mm{rr}}{ r} \nonumber \\
Q^{\theta}_\mm{vis}  & = &
\DD{1}{r^2} \DD{\partial}{\partial r} (r^2 T_\mm{r\theta})
 + \DD{1}{r \cos{\theta} }\DD{\partial  }{\partial \theta} (\cos{\theta} T_{\theta \theta})
 + {T_\mm{\phi\phi}}\tan{\theta} \\
Q^{\phi}_\mm{vis}  & = &
\DD{1}{r^2} \DD{\partial}{\partial r} (r^2 T_\mm{r\phi})
 + \DD{1}{r \cos{\theta} }\DD{\partial  }{\partial \theta} (r {\cos}^2{\theta}\, T_{\theta \varphi}), \nonumber
\end{eqnarray}
where
\begin{eqnarray}
T_\mm{rr}  & = &  2 \eta (\DD{\partial \Vpr}{\partial r} -
\DD{1}{3}(\DD{1}{r^2}\DD{\partial r^2 V_{r}}{\partial r}
 + \DD{1}{r \cos{\theta} }\DD{\partial  }{\partial \theta} (\cos{\theta} \,\Vpz))  \nonumber \\
T_\mm{\theta \theta} &=& 2 \eta (\DD{1}{r} \DD{\partial \Vpz}{\partial \theta}  + \DD{\Vpr}{r}
-\DD{1}{3}(\DD{1}{r^2}\DD{\partial r^2 V_{r}}{\partial r}
 + \DD{1}{r \cos{\theta} }\DD{\partial  }{\partial \theta} (\cos{\theta} \,\Vpz)))   \nonumber \\
T_\mm{\phi \phi} & = & 2 \eta (\DD{\Vpz}{r}\tan{\theta}  + \DD{\Vpr}{r}
-\DD{1}{3}(\DD{1}{r^2}\DD{\partial r^2 V_{r}}{\partial r}
 + \DD{1}{r \cos{\theta} }\DD{\partial  }{\partial \theta} (\cos{\theta} \,\Vpz)))   \\
T_\mm{\theta \phi} &=& \eta \DD{\cos{\theta}}{r} \DD{\partial}{\partial \theta} (\DD{\Vt}{\cos{\theta}})  \nonumber\\
T_\mm{r \phi} &=& \eta r \DD{\partial }{\partial r} (\DD{\Vt}{r})   \nonumber\\
T_\mm{r \theta} &=& \eta (r \DD{\partial}{\partial r} (\DD{\Vpz}{r})
 +  \DD{1}{r} \DD{\partial \Vpr}{\partial \theta}).     \nonumber
\end{eqnarray}
\subsection{The isotropic radiation transfer equation: The Kompaneets equation}

Compton up-scattering of soft photons is most efficient in
unsaturated Comptonization regions where the Compton-Y parameter
is of order unity. This parameter acquires large values in
optically thick media, and small values in the corona, implying
that  the corona-disk interaction region and/or the innermost
region of the disk are  most appropriate for this process to
operate efficiently. As a consequence, Comptonization in accretion
flows is intrinsically two-dimensional, and therefore requires a
multi-dimensional treatment.
\begin{table*}
\begin{tabular}{llll}
{Symbols:} \hspace*{-1.5cm} &&&\\ \hline
V &=       & $(\Vpr,\Vpz,\Vt)$                                     & velocity field  \\
B &=       & $(\Bpr,\Bpz,\Bt) = (\Bp,\Bt)$                         & magnetic field  \\
$\nabla$&= & $ (\DD{\partial }{\partial r}, \DD{1}{r }\DD{\partial  }{\partial \theta}) $  & gradient in spherical coordinates\\
$\nabla \cdot$&=&  $ \DD{1}{r^2} \DD{\partial }{r} r^2
    + \DD{1}{r \cos{\theta} }\DD{\partial  }{\partial \theta}  {\cos}{\theta}$ & divergence in spherical coordinates \\
$T^\mm{e,i}$ &=&   &electron and ion temperatures \\
$P^\mm{e,i}$ &=&  ${\cal R}_{\rm gas} \rho (T_i/\mu_{\rm i}
+ T_e/\mu_{\rm e}))$ &electron and ion pressure \\
${\cal E}^\mm{e,i}$ &=&  $P^{e,i}/(\gamma-1),$ & electron and ion internal energies  \\
$\kappa ^\mm{e,i}$ &=&  $ 7.8\times T^{3/2}_e,\, 3.2\times T^{3/2}_i$ & electron and ion conductivities  \\
$(m,n,\ell)$ &=  &  $\rho(\Vpr,r\,\Vpz,r\cos{\theta}\,\Vt)$   & momentum \\
$\nu(=\eta/\rho), \nu_\mm{mag}$ & &  & turbulent and magnetic diffusivities \\
$ \Phi$ & =  & $\Phi_\mm{HD} + \Phi_\mm{MHD}$  & HD and MHD turbulent dissipation (see MM). \\ \hline
\end{tabular}
\caption{Variables used and their definitions}
\end{table*}
So far, Comptonization has been considered under strong
assumptions that allow
 separation of variables and lead to the separation of the Kompaneets operator from
the radiative transfer equation. Here, the radiative intensity is
assumed to be time-independent, isotropic and the plasma is
isothermal. In this case, the generation and Comptonization  of
photons can be described by a second order differential equation
in the frequency space
 \cite{Iilarinov, Felten, Katz, Shapiro, HUJ3}.

Different accretion models display different spectra. Therefore,
it is essential to perform a diagnostic study to analyze their
consistency with observations. This however requires solving the
7D radiation transfer equation:
\begin{equation}
\DD{1}{c} \DD { \partial {I}}{\partial t} + n \cdot \nabla I = \kappa_{\nu}
\rho (S_{\nu}-I) - \sigma \rho I
           + \int{CI~d\acute{\Omega} d\acute{E}} + \varepsilon^\mathrm{mod}_{\nu},
\end{equation}
where $I=I(t,r,\theta,\varphi,\vartheta,\phi,\nu)$ is the
radiative intensity which depends on time t, the spherical
coordinates $(r,\theta,\varphi)$, two ordinates $(\vartheta,\phi)$
that determine the direction of the photons on the unit sphere,
and on the frequency $\nu$. $\kappa_{\nu}$ and $\sigma$ are the
absorption and scattering coefficients. $S_{\nu}$ is a source
function.
 $ I_\mathrm{int} \doteq  \int{CI~d\acute{\Omega} d\acute{E}}$ describes the
scattering of photons through electrons, and $C$ is the scattering
kernel. $\varepsilon^\mathrm{mod}_{\nu}$ is the modified
synchrotron emission.

To make the problem tractable, the following approximations have
been performed:
\begin{itemize}
\item The radiation field is axi-symmetric and isotropic, i.e., $\partial/\partial \varphi = 0$ and
      $ J = \DD{1}{2\pi} \int{I d \acute{\Omega}} \approx I.$
\item The source function is represented by the modified black body function, i.e.,
\begin{equation}
 S_{\nu} = B_{\nu}^\mathrm{mod} = \DD{2 B_{\nu}}{1 + \sqrt{1 + \DD{\sigma}{\kappa_{\nu}}}},
\end{equation}
where $B_{\nu}$ is the normal Planck function \cite[see][]{Rybiki}.
\item The thermal energy of the electrons is far below its corresponding rest mass energy,
     i.e., $\epsilon = \DD{kT}{m_\mathrm{e}c^2} \ll 1,$ and $\DD{h \nu}{m_\mathrm{e}c^2} \ll 1.$
\end{itemize}
Using the last approximation, $ I_\mathrm{int}$ can be expanded up
to second order in $\epsilon$  which reduces it to the so-called
Kompaneets operator \cite{Payne}:
 \begin{equation}
 I_\mathrm{int} \LRA {\cal K}_{\nu}  = -\DD{\nu}{m_\mathrm{e}c^2}\DD{\partial}{\partial \nu}(4kT-h\nu)I +
 \DD{kT\nu}{m_\mathrm{e}c^2}\DD{\D^2}{\partial {\nu}^2} (\nu I).
\end{equation}
In this case, the radiative transfer equation with respect to a
rest frame of reference reads:
\[\DD{1}{c}[\DD { \partial {E_{\nu}}}{\partial t}
  + \nabla  \cdot V E_{\nu}] =
   -\lambda_{\nu} (\nabla  \cdot V){E_{\nu}}  +
   \nabla  \cdot [\DD{\lambda_{\nu}}{\chi_{\nu}}\nabla E_{\nu}]\]
\begin{equation}
\hspace*{2.0cm} + \kappa_{\nu} \rho (S_{\nu}-E_{\nu}) + {\cal
K}_{\nu} + \varepsilon^\mathrm{mod}_{\nu}, \label{Eq_RTE}
\end{equation}
where $E_{\nu}=\DD{4\pi}{c}J= E(t,r,\theta,\nu)$, $\chi_{\nu} =
\rho(\kappa_{\nu}+\sigma)$, $\lambda_{\nu}$ is the
flux limited diffusion coefficient \cite{Levermore},
which forces the radiative flux to adopt the correct form in
optically thin and thick regions, i.e.,
\begin{equation}
{\nabla  \cdot [{\lambda_{\nu}}{}\nabla E_{\nu}] =
        \left\{ \begin{array} {c@{\quad \quad}l}
  \nabla \cdot [\DD{1}{3\chi_{\nu}}\nabla E_{\nu}]  & {\sf if} \hspace*{0.75cm} \tau \gg 1  \\
  \nabla \cdot nE_{\nu}  & {\sf if} \hspace*{0.75cm} \tau \ll 1.
        \end{array} \right. }
\end{equation}
$\lambda_{\nu}$ may provide a smooth matching between these two
extreme regimes.  The above two different behaviour of the
operator can be combined as follows:
\begin{equation}
\nabla  \cdot {\lambda_{\nu}}{}\nabla_{\nu} E_{\nu} \hookrightarrow
\nabla  \cdot \eta_\mathrm{r}\nabla E_{\nu},
\end{equation}
where $\eta_\mathrm{r} = (1-\alpha) \DD{\nabla E_{\nu}}{ |\nabla
E_{\nu}|} + \alpha \DD{1}{3\chi} ,$  $\alpha =
e^{-R_\mathrm{FLD}}$, and
 $R_\mathrm{FLD} = \nabla E/\rho(\kappa_{\nu}S_{\nu}+\sigma E_{\nu})$ \cite{HUJ0}.

$\varepsilon_{\nu}^\mathrm{mod}$ in Eq. (15) corresponds to the
modified synchrotron emission of photons by relativistic electrons
gyrating around magnetic field lines, which reads:
\[
  \varepsilon_{\nu}^{mod} = \xi~\varepsilon_{\nu} + (1-\xi)\varepsilon^{BB}_{\nu},
\]
where $\xi (\doteq e^{-({{\nu}_\mathrm{c}}/{\nu})^2})$ is a switch
on/off operator which bridges optically thin and thick media to
synchrotron radiation, and ${\nu}_{c}$ is a critical frequency
(see below).

An appropriate  approximation for  $\varepsilon_{\nu}$ in
optically thin medium reads \cite{Mahadevan}:
\begin{equation}
\varepsilon_{\nu} = 2.73\times 10^{-5} \DD{\rho \nu
}{K_{2}(1/\theta_\mathrm{e})}{\tilde{\cal I}}(\nu,B,\Theta)~ ~{\rm
ergs~cm^{-3}~s^{-1}~ {\rm Hz}^{-1}},
\end{equation}
where $K_{2}$ is the Bessel function of the second kind  and
$\tilde{\cal I} = \DD{4.05}{\zeta^{1/6}}(1 + \DD{0.4}{\zeta^{1/4}}
+ \DD{0.53}{\zeta^{1/2}})
 e^{-1.89 \zeta^{1/3}}.$\\
Here $\zeta = 2.38\times 10^{-7}
({\nu}/{B{\theta}^{2}_\mathrm{e}})$  and
 $\theta_\mathrm{e} = {k T_\mathrm{e}}/{m_\mathrm{e} c^2}.$\\
Below a certain critical frequency $\nu_\mathrm{c}$, the media
become self-absorbing to synchrotron emission. In this case, $
\varepsilon_{\nu}^\mathrm{mod} \approx
\varepsilon^\mathrm{BB}_{\nu} ={2\pi}\DD{\nu^2}{c^2}kT. $ To find
$\nu_\mathrm{c}$, we use the  local non-linear Newton iteration
procedure applied
 to the equation
\begin{equation}
   \int_{V}{\varepsilon_{\nu}dV} = \int_{S}{\varepsilon^\mathrm{BB}_{\nu}} dS.
\end{equation}
Having obtained $\nu_\mathrm{c}$, the switch on/off operator $\xi$
can then be constructed.
\begin{figure*}
\centering {\hspace*{-0.2cm}
\includegraphics*[width=13.75cm, bb=5 160 587 835,clip]
{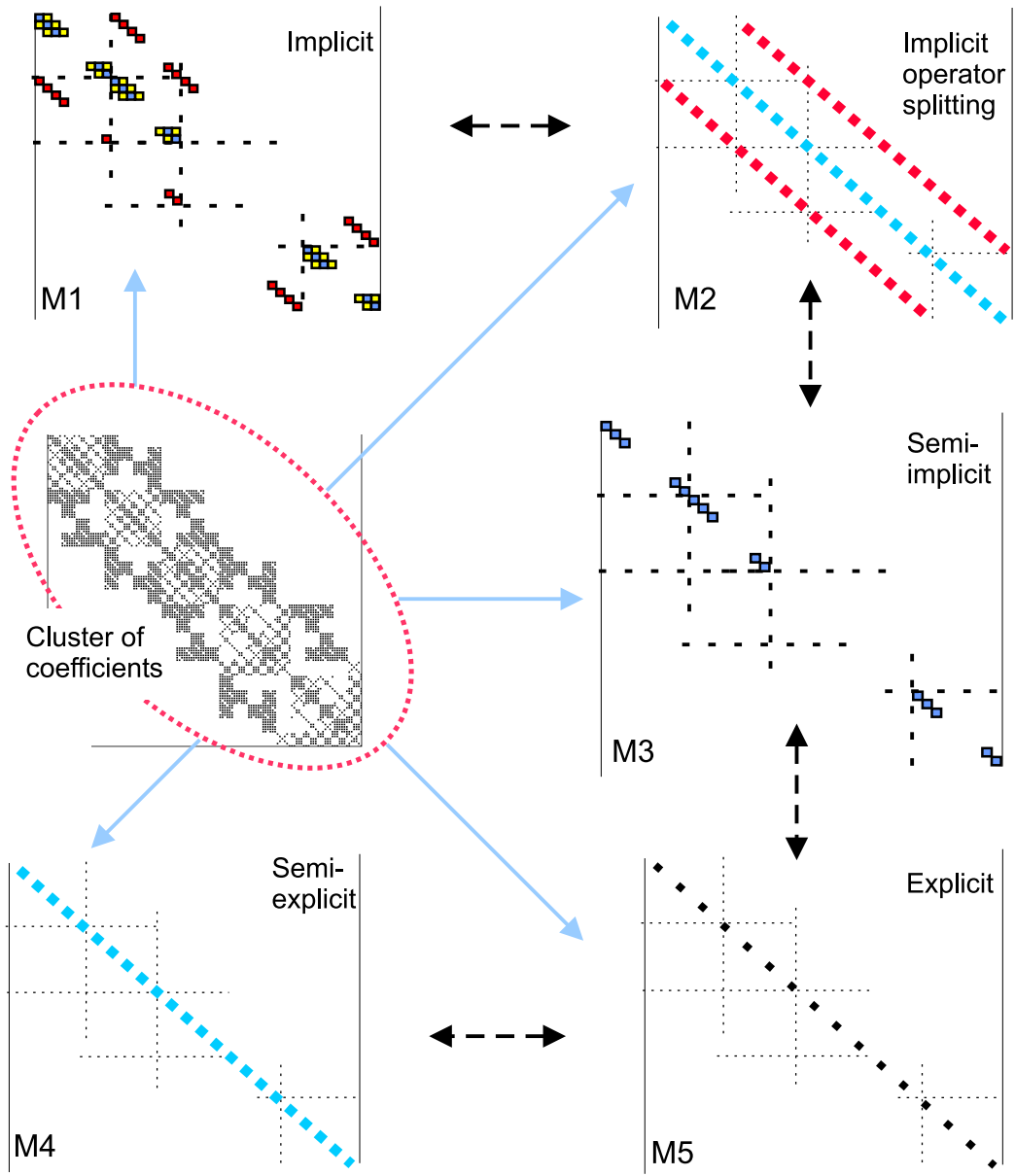} } \caption{ A schematic description of the
hierarchical solution method. A cluster of coefficients is
computed in the first stage, and a matrix-generator is created
that allows using various solution procedures ranging from purely
explicit to fully implicit. Interchange between solution methods
is possible, as modifying, adding or removing entries is directly
maintainable.
   }
\end{figure*}
\section{Solution methods }
\subsection{Solving the radiative MHD equations}
The set of equations in conservative form  may be written in the following vector form:
\beq
 \DD{\partial \vec{q}}{\partial t} + L_\mm{r,rr} \vec{F} +   L_\mm{\theta,\theta\theta} \vec{G}  = \vec{f},
\eeq where $F$ and $G$  are fluxes of $q$, and $ L_\mm{r,rr},\,
L_\mm{\theta,\theta\theta}$ are first and second order transport operators
 that describe advection-diffusion
of  the vector variables $\vec{q}$ in $\mm{r}$ and $\theta$  directions.
$\vec{f}$ corresponds
to the vector of source functions.\\
\begin{figure*}
\begin{center}
\hspace*{-2.5cm}
\includegraphics*[angle=90, width=7.5cm, bb=0 0 313 450,clip ]{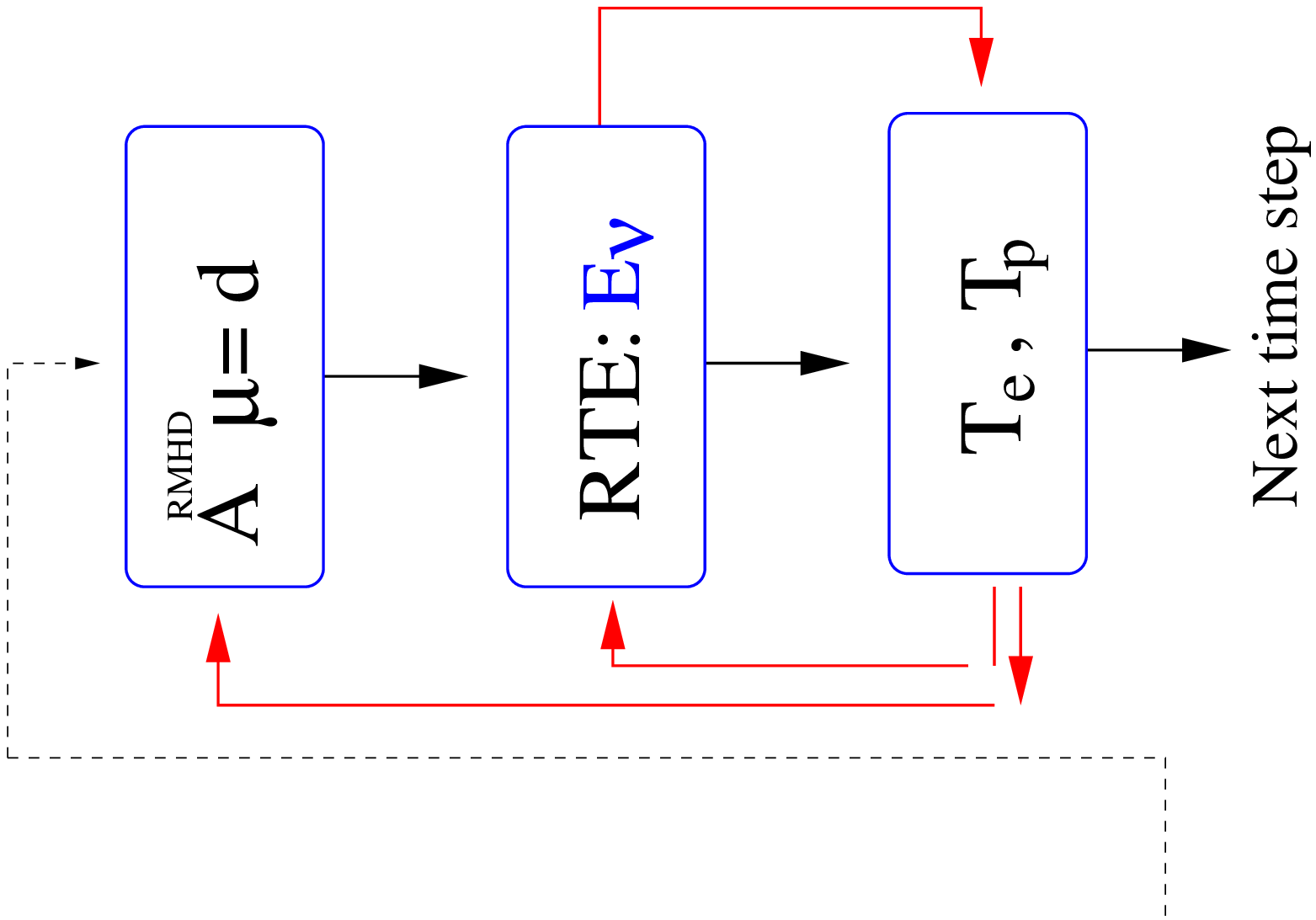} \\
\includegraphics*[angle=90, width=11.5cm, bb=0 0 498 460,clip ]{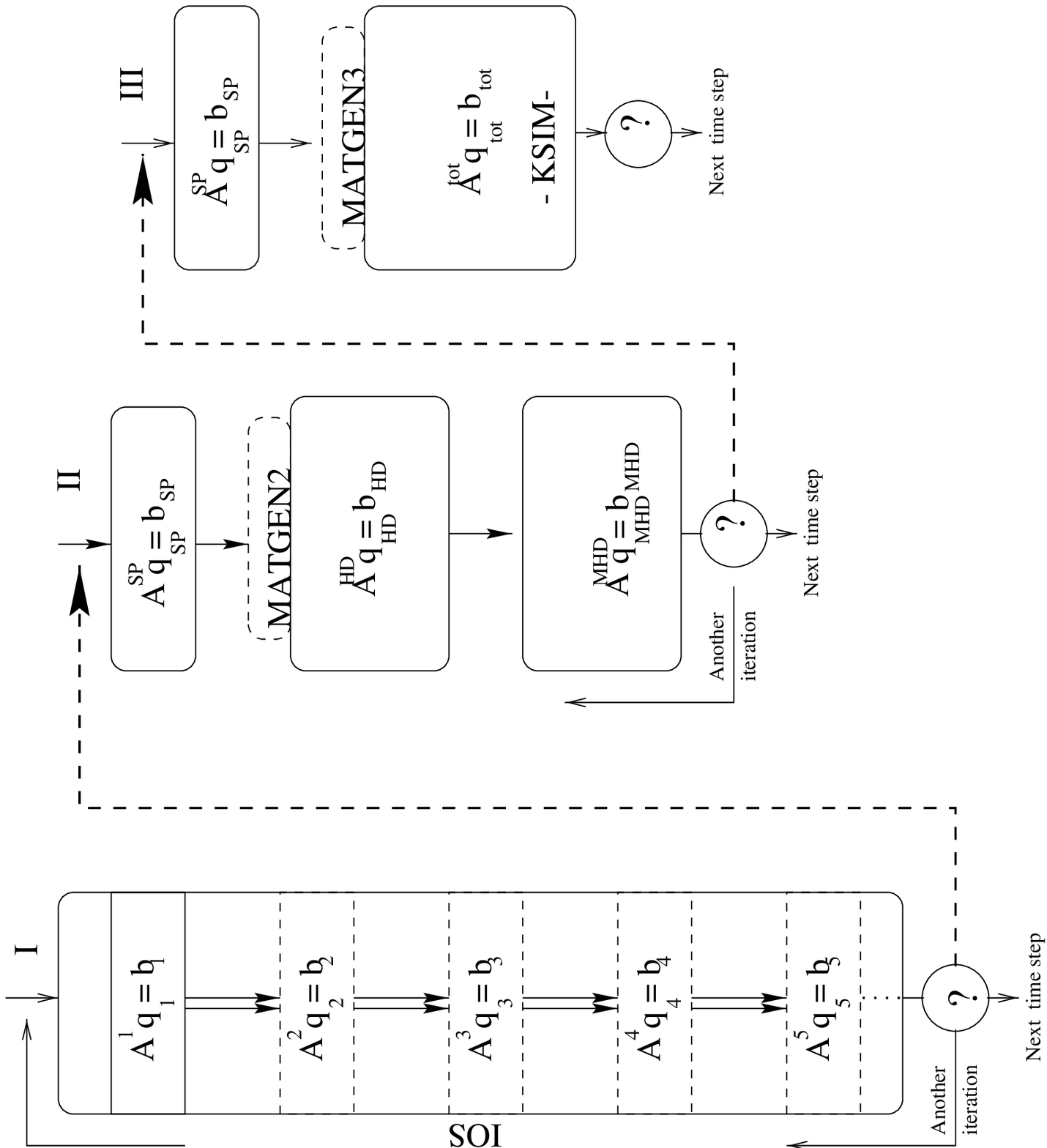}
\hspace*{-0.5cm}
\end{center}
\caption[ ]{ \small  A schematic description of the hierarchical
algorithm for solving the radiative MHD equations. Stage I
corresponds to the implicit operator splitting approach (IOS),
which is most appropriate for following the early time-dependent
phases of the flow. The solution obtained can then be used as
initial condition for Stage II, where   the hydro-equations are
solved as a single coupled system, followed by the  magneto
component, which is again solved as a single coupled system. Here,
high spatial and temporal accuracies in combination with the
prolongation/restriction strategy may be used. Similarly,  the
solution obtained in this stage may be used as starting solutions
for Stage III, where  steady
        solutions for the fully coupled set of equations consisting of
the zero moment of the radiation field and the MHD equations are
sought. In this stage, pre-conditioned Krylov sub-iterative
methods are considered to be robust and efficient. The very last
stage, Stage IV,  corresponds to the case where solutions for the
internal energy equations   weakly coupled with the 5D radiative
transfer equation are sought. }
\end{figure*}
Adopting a five star staggered grid discretization, it is easy to
verify that at each grid point the Eq. (23) acquires the following
block matrix equation:
\[
\DD{{\delta q}_\mm{j,k}}{\delta t} + \underline{S}^\mm{r}{\delta
q}_\mm{j-1,k} + {D}^\mm{r}{\delta q}_\mm{j,k}
                + \overline{S}^\mm{r}{\delta q}_\mm{j+1,k} \]
\beq
 + \underline{S}^\mm{\theta}{\delta q}_\mm{j,k-1} + {D}^\mm{\theta}{\delta q}_\mm{j,k}
                + \overline{S}^\mm{\theta}{\delta q}_\mm{j,k+1}
= RHS^\mm{n}_\mm{j,k},
\eeq
 where the subscripts ``j'' and ``k'' denote
the grid-numbering in the $\mm{r}$ and $\theta$ directions,
respectively, and $RHS^\mm{n}=[\vec{f}- L_\mm{r,rr} \vec{F} -
L_\mm{\theta,\theta\theta} \vec{G}]^\mm{n}$. Underlines
(overlines) mark the sub-diagonal (super-diagonal) block matrices
in the corresponding directions, and
 ${D}^\mm{r,\theta}$ are the
diagonal block matrices.\\
To outline the directional dependence of the block matrices, we
re-write Eq. 24 in a more compact form: \beq
 \begin{array}{lll}
 & {\hspace*{0.3cm}}\overline{S}^\mm{\theta}{\delta q}_\mm{j,k+1} & \\
+ \underline{S}^\mm{r}{\delta q}_\mm{j-1,k} & +
{D}_\mm{mod}{\delta q}_\mm{j,k} &
  + \overline{S}^\mm{r}{\delta q}_\mm{j+1,k}   = RHS^\mm{n}_\mm{j,k} \\
& + \underline{S}^\mm{\theta}{\delta q}_\mm{j,k-1}, &
\end{array}
\eeq
  where \({D}_\mm{mod} = {{\delta q}_\mm{j,k}}/{\delta t} +
{D}^\mm{x}+ {D}^\mm{y}.\) Eq. (25) gives rise to at least four
different types of solution procedures: \ben
\item Classical explicit methods are very special cases in which
the sub- and super-diagonal block matrices together with
         ${D}^\mm{x}$ and ${D}^\mm{y}$ are neglected. The only  matrix to be retained here
        is $ (1/{\delta t})\,\times\,$(the identity matrix), i.e.,
        the first term on the LHS of Eq. 24. This yields the vector equation (see M5/Fig. 2):
    \beq
        [\DD{I}{\delta t}]{\delta q}_\mm{j,k}  = RHS^\mm{n}_\mm{j,k}.
    \eeq
\item Semi-explicit methods are obtained by preserving the diagonal
entries,
      $d_\mm{j,k},$ of the block diagonal matrix ${D}_\mm{mod}$ (see M4/Fig. 2). This method has been
      verified to be numerically stable even when large Courant-Friedrich-Levy (CFL) numbers
      are used. In particular, this method is absolutely stable if the flow is
      viscous-dominated.
\item Semi-implicit methods are recovered when neglecting the sub- and super-diagonal
  block matrices only, but retaining the block diagonal matrices (see M3/Fig. 2). In this case
      the matrix equation reads:
    \beq
       {D}_\mm{mod}{\delta q}_\mm{j,k} = RHS^\mm{n}_\mm{j,k}.
    \eeq
       We note that inverting ${D}_\mm{mod}$ is a straightforward procedure, which can be maintained
       analytically or numerically.
\item A fully implicit solution procedure requires retaining
       all the  block matrices on the LHS of Eq. 25. This yields a global matrix that
       is highly sparse (M1/Fig.2).  In this case, the
      ``Approximate Factorization Method''  \cite[-AFM:][]{Beam} and the
     ``Line Gauss-Seidel Relaxation Method'' \cite[-LGS:][]{MacCormack}
       are considered to be efficient preconditionings for the set of radiative MHD-equations.
\een
In the case that only stationary solutions are sought, convergence to steady state can
be accelerated by adopting the so called the ``Residual Smoothing Method'' 
\cite[see][and the references therein]{HUJ8}.\\
This method is based on associating a time step size with the local CFL-number at
each grid point.  While this strategy is efficient
    at providing quasi-stationary solutions within a reasonable number of iterations,  it
is incapable at providing
physically meaningful time scales for features that possess
quasi-stationary behaviour. Here we suggest to use the obtained
quasi-stationary solutions as initial configuration and re-start
the calculations using a uniform and physically relevant time steps.
\begin{figure*}
\begin{center}
{\hspace*{-0.5cm}
\includegraphics*[width=12.5cm,bb=40 175 385 600,clip]{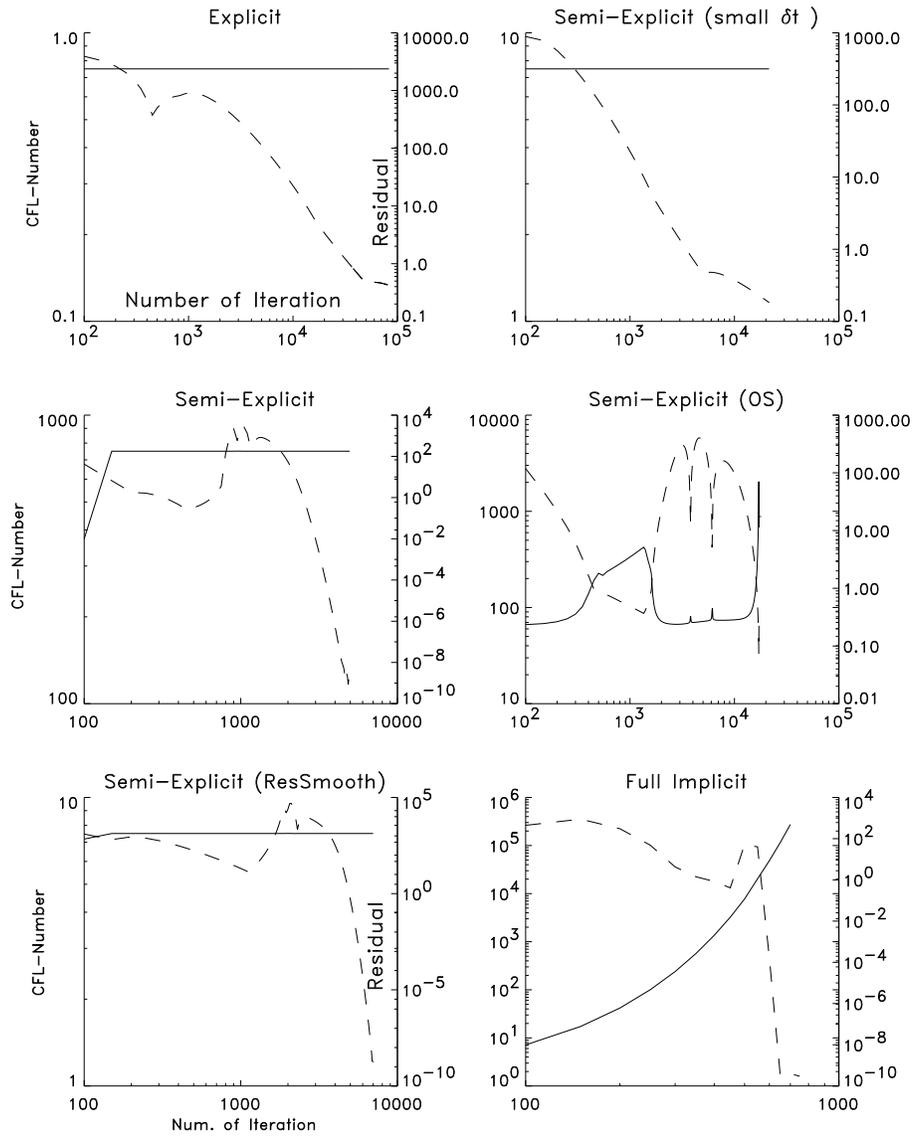}
}
\end{center}
{\vspace*{-0.0cm}} \caption [ ] { The problem of free-fall of gas onto
a Schwarzschild black hole. The evolution of the CFL-number and the
residual versus number of iterations are shown, using different solution procedures.
 The solution
methods are: normal explicit (top/left), semi-explicit
(middle/left), semi-explicit in combination with the residual
smoothing strategy(bottom/left), semi-explicit using moderate
CFL-numbers (top/right), semi-explicit method in which the time
step size is taken to be  a function of the maximum residual
(middle/right), and finally the fully implicit method
(bottom/right). The different forms of the
semi-explicit method used here are stable and converges to the
stationary solution, though at remarkably different rates.
  }
\end{figure*}
\begin{figure}
\begin{center}
{\hspace*{-0.5cm}
\includegraphics*[width=6.25cm,bb=45 33 275 740,clip]{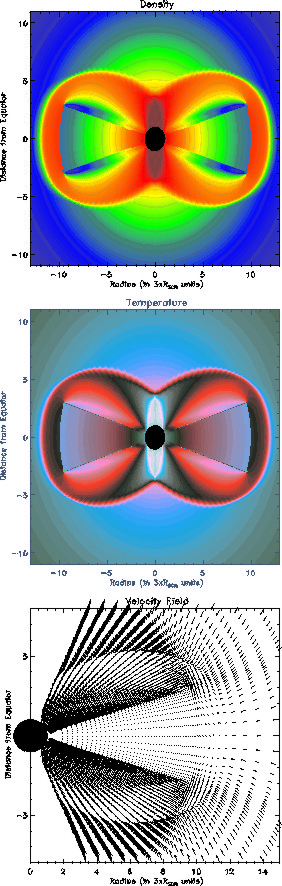}
}
\end{center}
{\vspace*{-0.0cm}} \caption [ ] {  Free-fall of gas onto a black
hole surrounded by a static cold disk. Top: the density distribution
 (red: large values, blue: low values, green: intermediate values).
  Middle: the temperature distribution (red: large values, blue: low
  values, gray: intermediate values). The curved shock front,
  where the temperature attains maxima is obvious.
  Bottom: the distribution of the velocity field is shown.
  }
\end{figure}
\subsection{The 5D axi-symmetric RT equation: method of solution}
Let ${\cal L}E=0$ be the equivalent operator form of Equation 18
in the continuous space $\Omega_C$. $\cal{L} E$ consists  of
several terms, each of which requires a careful and different
representation in the finite discretization space $\Omega_h$ which
is defined as
$[t_{1},t_{2},...,t_{N}]\otimes[r_{1},r_{2},...,r_{J}]
\otimes[\theta_{1},\theta_{2},...,\theta_{K}]\otimes[\nu_{1},\nu_{2},...,\nu_{M}]$.
$\rm{[t],~[r],~[\theta]}$ and $[\nu]$ correspond to time, radius
(spherical), latitude, and to the frequency intervals,
respectively.

In most astrophysical problems, radiative effects occur on
relatively short time scales compared to the hydro- or
magneto-hydrodynamical ones, for which the use of unconditionally
implicit numerical solvers is essential. This requires however
that  all terms of Eq. (18)  should be evaluated on the new
time-level.  The discretization used should  assure that the
resulting Jacobian $A_\mathrm{r\theta\nu} = \partial {\cal L} E/\partial E$
is diagonally dominant. Therefore, the following procedures are employed.\\
\bit
\item The advection term $\nabla \cdot V E_{\nu} $ is discretized
 using a second order up-winding. \\
\item The second order diffusion term  $\nabla  \cdot [{\lambda_{\nu}}\nabla E_{\nu}]$
   is discretized  using second order central-difference scheme on a staggered grid \\
\item ${\cal K}_{\nu}$ contains advection and diffusion
terms in the frequency space. Here up-winding discretization in
the  frequency space is used. \eit Combining the contributions  of
all terms of Eq. (18), we obtain at each grid point the following
equation:
\[ {  \underline{S}^\mathrm{r} E^\mathrm{new}_\mathrm{j-1,k,m} + \overline{S}^\mathrm{r} E^\mathrm{new}_\mathrm{j+1,k,m}
  + \underline{S}^{\theta} E^\mathrm{new}_\mathrm{j,k-1,m} }\]
\[ {+ \overline{S}^{\theta} E^\mathrm{new}_\mathrm{j,k+1,m}  }
 {+ \underline{S}^{\nu} E^\mathrm{new}_\mathrm{j,k,m-1} + \overline{S}^{\nu}  E^\mathrm{new}_\mathrm{j,k,m+1}} \]
\begin{equation}
{ + (D^\mathrm{r} + D^{\theta}   + D^{\nu})
E^\mathrm{new}_\mathrm{j,k,m} = RHS},
\end{equation}
where {{$\underline{ S}^\mathrm{r}          = \partial {\cal L}
E_{\nu}/\partial E_\mathrm{j-1,k,m}$}},
       {{$\overline{ S}^\mathrm{r}           = \partial {\cal L} E_{\nu}/\partial E_\mathrm{j+1,k,m}$}},
       {{${\rm D^\mathrm{r} + \rm D^\theta   +  D^\nu} = \partial {\cal L} E_{\nu}/\partial \rm E_\mathrm{j,k,m}$}},
       {{$\underline{ S}^{\theta}     = \partial {\cal L} E_{\nu}/\partial E_\mathrm{j,k-1,m}$}},
       {{$\overline{S}^{\theta}      = \partial {\cal L} E_{\nu}/\partial  E_\mathrm{j,k+1,m}$}},
       {{$\overline{ S}^{\nu}         = \partial {\cal L} E_{\nu}/\partial  E_\mathrm{j,k,m+1}$}}
           and
       {{$\underline{ S}^{\nu}        = \partial {\cal L} E_{\nu}/\partial  E_\mathrm{j,k,m-1}$}}.
The terms $\underline{ S}^\mathrm{r},  D^\mathrm{r},$ and $
\overline{ S}^\mathrm{r}$  correspond to
 the sub-diagonal, diagonal
and super-diagonal entries of the Jacobian
$A_\mathrm{r\theta\nu}$ in the radial direction respectively.
 A similar description applies
 to the $\theta-$ and $\nu-$directions.

 Thus, solving the equation at all grid points, is equivalent to
  solve matrix equation:
 $ A_\mathrm{r\theta\nu}  E^\mathrm{new} =  E^\mathrm{old}$, or simply, $ Aq=b$.

This matrix is highly sparse, and pre-conditionings such as the
Alternating Direction Implicit (ADI) and the Approximate
Factorization Methods (AFM) are considered to be efficient.
However,  ADI is not appropriate for searching steady solution in
three or more dimensions, as it is numerically unstable in high
dimensions \cite{Fletcher}. Alternatively, we have tried the AFM
 as a pre-conditioner. However,  it turns out that the AFM converges slower
than our favorite iterative method: `Black-White-Brown' line
Gauss-Seidel method \cite[henceforth BWB-LGS, see][for further details]{Hirsch}. The latter method preserves the diagonal
dominance of ${A}$, and hence converges faster than AFM. It should
be noted that the  line Gauss-Seidel method in its classical form
is not appropriate for vector and parallel machines, mainly
because the vector-length is proportional to the number of
unknowns in one direction. A reasonable way to extend the
vector-length is to solve for all unknowns located on
even-numbered grid points, and subsequently on odd-numbered grid
points. The resulting vector-length in this case is proportional
to the number of unknowns in the plane under consideration, and
therefore  enabling enhancement efficiency when using vector or
parallel machines.

More specifically, in each plane we perform two sweeps: in the
first sweep we consider the unknowns in the  ${r-\theta}$ plane,
i.e., we solve the system of equations:
\[  \underline{S}^\mathrm{r} \delta E^\mathrm{new}_\mathrm{j-1,k,m}
     +  (D^\mathrm{r}+D^{\theta}+D^{\nu}) \delta E^\mathrm{new}_\mathrm{j,k,m}
   +  \overline{S}^\mathrm{r} \delta E^\mathrm{new}_\mathrm{j+1,k,m} = RHS, \]
 where $j=1\rightarrow J$ and k runs over odd-numbered rows. In the second sweep, we solve:
\[   \underline{S}^\mathrm{r} \delta E^\mathrm{new}_\mathrm{j-1,k,m}
     +  (D^\mathrm{r}+D^{\theta}+D^{\nu}) \delta E^\mathrm{new}_\mathrm{j,k,m}
     +  \overline{S}^\mathrm{r} \delta E^\mathrm{new}_\mathrm{j+1,k,m}\]
\[ = RHS
   + \underline{S}^{\theta} \delta E^\mathrm{new}_\mathrm{j,k-1,m} + \overline{S}^{\theta} \delta
  E^\mathrm{new}_\mathrm{j,k+1,m}, \]
where $j=1\rightarrow J$ and k here runs over even-numbered rows.
Therefore, we actually perform 6-inversion procedures per each
time step. Here the 3-dimensional problem is replaced by three
one-dimensional problems that are solved  iteratively to recover
the solution of the original problem. The method is relatively
efficient, as the overall number of arithmetic operations scales
linearly with the number of grid points (${\rm \sim 6\times
9\times N})$.
\begin{figure}[htb]
\begin{center}
{\hspace*{-0.5cm}
\includegraphics*[width=12.5cm, bb=55 390 500 710,clip]
{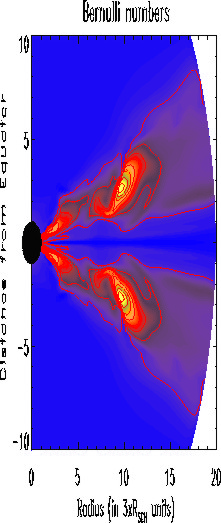} }
\end{center}
{\vspace*{-0.4cm}} \caption [ ] {  \small The distribution of the Bernoulli number
    of accretion flows around a supermassive black hole.
The Bernoulli number characterizes the energies  of the flow in different regions.
Gravitationally bound flows have negative total energy, whereas flows of positive total
energy are gravitationally unbound, and potentially should expand to infinity.
 In this figure, the decrease of the Bernoulli number
from large to low positive values is represented by yellow, green and
red colours, whereas  the blue colour corresponds to negative values.  Obviously,
  gravitationally unbound blobs are formed in the vicinity of the black hole, which
 thereafter collimate under the action of magnetic fields to form the observed
 highly collimated jets.
    }
\end{figure}
\begin{figure}[htb]
\begin{center}
{\hspace*{0.2cm}
\includegraphics*[width=7.5cm, bb=35 482 272 631,clip]
{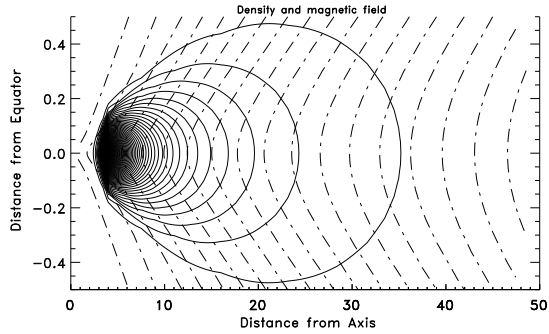} }
\end{center}
{\vspace*{-0.4cm}} \caption [ ] {The initial distribution of the density
in an accretion disk ($\Mdot=0.1 \Mdot_\mathrm{Edd}$)  around a Schwarzschild black hole
overlied by coronal plasmas (solid lines).  The dashed lines correspond to the
magnetic field lines threading the disk and the corona.
The distance is given in  units of $2.75~R_\mathrm{Sch}$, where $R_\mathrm{Sch}$ is
the Schwarzschild radius.
  }
\label{FigLD3}
\end{figure}
\begin{figure}[htb]
\centering {\hspace*{0.2cm}
\includegraphics*[width=7.5cm, bb=35 500 275 740,clip]
{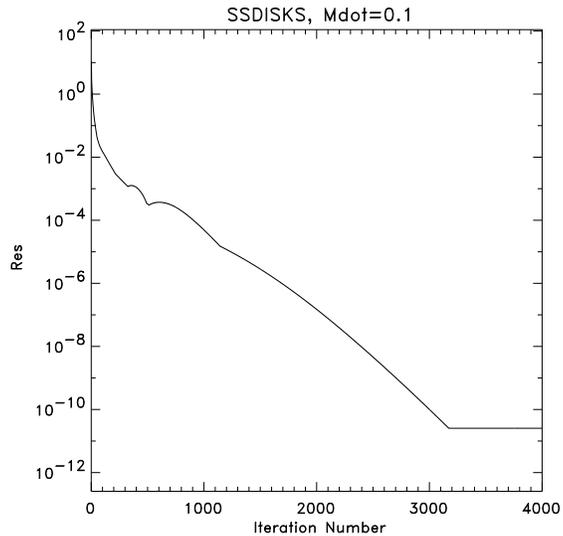} } \caption{The evolution of the residual in the maximum
norm versus the number of iteration. The adopted density and temperature profiles
               correspond to standard accretion disks (see Fig. 7).  }
\end{figure}

\begin{figure}
\begin{center}
{\hspace*{-0.5cm}
\includegraphics*[width=11.0cm, bb=37 40 535 316,clip]{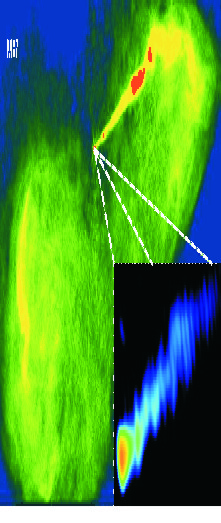}\\
\includegraphics*[width=11.0cm, bb=53 509 512 738,clip]{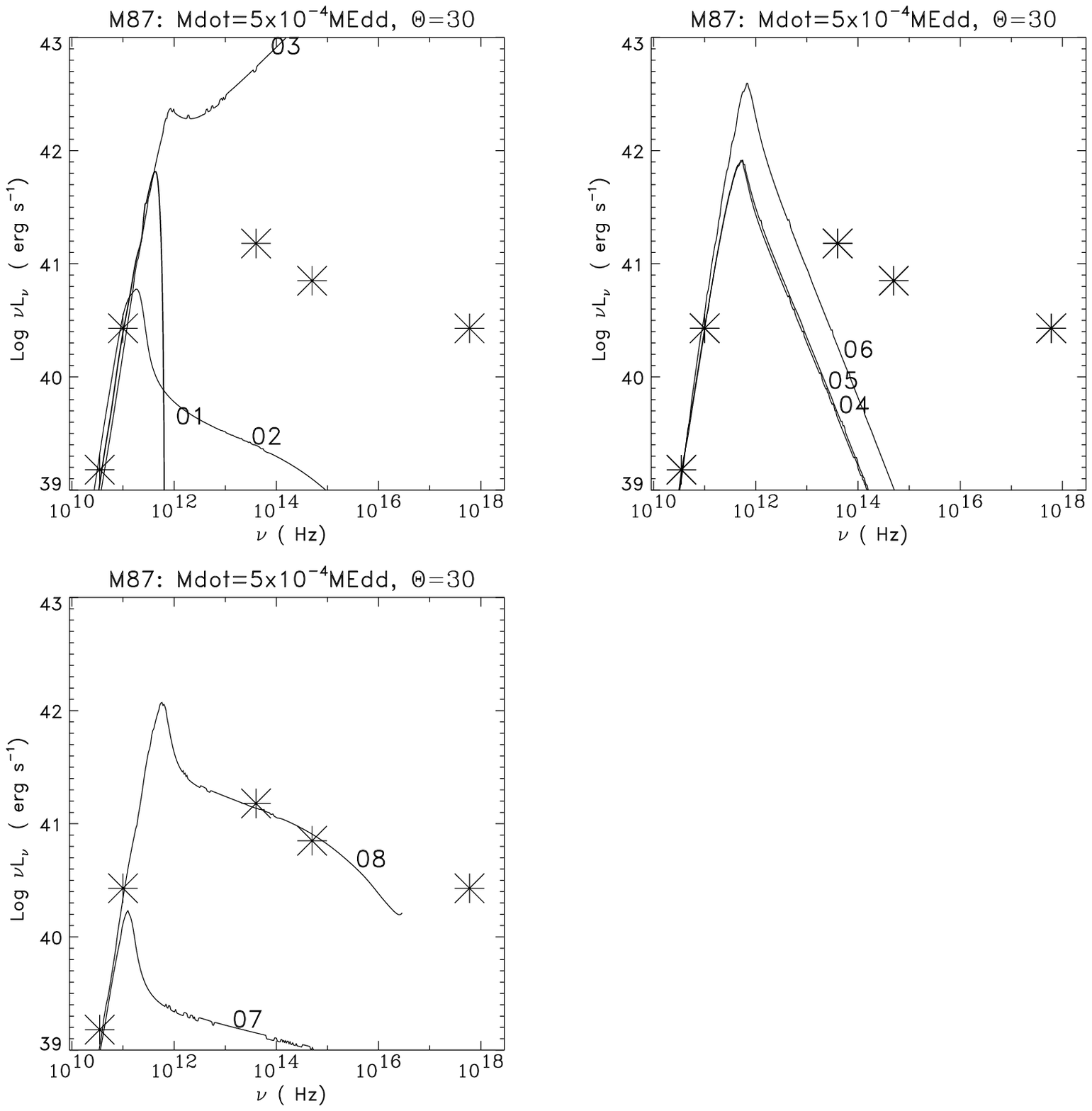} \\
\includegraphics*[width=6.0cm, bb=53 270 270 505,clip]{Fig9b.ps}
\vspace*{-0.75cm}
}
\end{center}
\caption [ ] { \small  A VLA image shows the active central engine of the giant
     elliptical galaxy M87 (top), and a NRAO radio image of the jet apparently emanating from
     within 100 gravitational radii. Solid lines correspond to calculated profiles and
   the asterisks to observational data.
   The profiles 01 to 06 show the spectral energy distribution calculated using
  different magnetic field strengths, or  different truncation radii, or high/low
  corona temperatures.
   In particular, the profile 07 corresponds to a model in which the toroidal magnetic
  field is set to vanish artificially, whereas the poloidal magnetic field is set to be in
   equipartition with the thermal energy of the electrons.
   The profile 08 is similar to 07, except that the toroidal magnetic field  is
   allowed to develop and reach values beyond equipartition with respect to the
   thermal energy of the electrons in the transition layer between the disk and
   the overlying corona.
   The above  spectral energy distribution has been obtained by solving
   the radiative transfer equation in 5-dimensions, taking into account the Kompaneets
   operator for consistently modelling Comptonization.  400 non-linearly distributed
    frequency points have been used to cover the frequency-space, and
    $125\times 40$ finite volume cells to cover the spatial domain of the calculation. }
\end{figure}
\section{Validation and preliminary tests}
\subsection{Free-fall of plasma onto a Schwarzschild black hole}
      A centrifugally-unsupported gas around a spinless black hole
      is gravitationally bound, and therefore should fall-freely onto
      the black hole, provided that no other external forces oppose gravity.
      In this case, the radial distributions of the density and velocity
    far from the event horizon obey the power laws:
$r^{-3/2}$ and $r^{-1/2}$, respectively.  \\
This physical problem is relevant for testing the flexibility of
the hierarchical scenario at adopting various solution methods,
and to test their capability to capture steady, oscillation-free
and advection-dominated flows, even when
a strongly stretched mesh distribution is used.\\
The equations to be solved in this problem are the continuity, the
radial and horizontal momentum equations, and the internal energy
equation. The flow is assumed to be inviscid and adiabatic ($\gamma =5/3$).
 The equations have been solved using a
first order accurate advection scheme both in space and time.   In
carrying out these calculations, the following conditions/inputs have
been taken into account:
\bit
    \item The central object is  a one  solar-mass
    and non-rotating black hole.
    \item { The outer boundary is  100 times larger than the
    the inner radius, i.e.,  $R_\mathrm{out}= 100\times
    R_\mathrm{in}$, where $R_\mathrm{in}$ is taken to be the radius of the last stable
    orbit\footnote{$R_\mathrm{LS} = 3\times R_\mathrm{S} = 6\times R_\mathrm{g}$, where $R_\mathrm{S}$ and
     $R_\mathrm{g}$ are the Schwarzschild and
    gravitational radii, respectively.} $R_\mathrm{LS}$ . To first order in $V/c$, the flow at this radius
    can be still treated as non-relativistic, though the error can be as large as 30\%.    }
\item Along the outer boundary,  the density and temperature of the gas
     assume uniform distributions, and flow across
     this boundary with the free-fall velocity. Symmetry boundary
    conditions along the equator, and asymmetry boundary conditions along the
     axis of rotation have been imposed. Along the inner boundary,
     we have imposed non-reflecting and outflow conditions.
     { This means that  up-stream conditions are imposed, which forbid
     information exterior to the boundary to penetrate into the domain of calculations.
     In particular, the actual values of the density, temperature and momentum in the ghost zone  r
     are erased and replaced by the corresponding values
     in the last zone, i.e, the zone between $R_\mathrm{in}$ and $R_\mathrm{in} + \Delta R$.
     In the case that second order viscous operators are considered, care has been taken to
     assure that their first order derivatives across $R_\mathrm{in}$ are vanished.  }
\eit
The above set of equations are solved in the first quadrant $[1\le
r \le 100]\times [0\le \theta \le \pi/2]$, where 200 strongly
stretched finite volume cells in the radial direction and 60 in
the horizontal direction
are used.\\
In  Fig. 4, we show  the evolutions of the CFL-number and the
residual as function of the number of iteration which has been
obtained using various  numerical approaches. The results show
that the convergence of the explicit and  semi-explicit methods
are rather slow when a relatively small time step size is used.
This implies that  the amplitude-limited  oscillations are
strongly time-dependent that may result from geometric
compression. Indeed, these perturbations disappear, when
relatively large
time-step sizes are used (see Fig. 4,  bottom/right). \\
In addition, the semi-explicit solver has been tested in
combination with the residual smoothing strategy. As expected,
this approach   accelerates  the convergence considerably (Fig. 4:
compare the plots bottom/left with  the top/right).

In most of the cases considered here, the time-step size is set to
increase in a well-prescribed manner and independent of the
residual.  However, determining  the size of the time step  from
the residual directly did not provide
satisfactory convergence histories (Fig. 4, middle/right). \\
The results obtained here indicate that the semi-explicit method
is stable  and can be applied to search for stationary
solutions using large time steps, or equivalently,  CFL-numbers
that are significantly larger than unity (Fig. 4, middle/left).
\subsection{Shock formation around black holes}
Similar to the forward facing step in CFD, a cold and dense disk
has been placed in the innermost equatorial region: $[1\le r\le
10]\times [-0.3 \le \theta \le 0.3] .$ We use the same parameters,
initial and boundary condition as in the previous flow problem. A
vanishing in- and out-flow conditions have been imposed at the
boundaries of this disk. The gas surrounding the disk is taken to
be inviscid, thin, hot and non-rotating. Thus, the flow
configuration is similar to the forward facing step problem
usually used for test calculations in CFD. The disk here serves as
a barrier that forbids the gas from freely falling onto the black
hole, and instead, it forms a curved shock front around the cold
disk. The purpose of this test is mainly to examine the capability
of the hierarchical scenario at employing the semi-explicit method
adequately and enables capturing steady solution governed by
strong shocks.
 In solving
the HD-equations, an advection scheme  of third order spatial
accuracy and first order accurate in time has been used. The
domain of calculation is sub-divided into 200 strongly-stretched
finite volume cells in the radial direction and 60 in horizontal
direction. In Fig. 5 the configuration of the steady distributions
of the density, temperature and the velocity field are shown.
Similar to the calculation in the previous sub-section, the
results indicate that the method employed is stable and converges
to the sought steady solution even when a CFL-number of order 200
is used. However, the method converges relatively slowly compared
to the implicit operator splitting approach, where steady
solutions have been obtained after one thousand iterations only.
\subsection{Formation and acceleration of proton-dominated jets in active galaxies}
      To study the mechanisms underlying jet formation around black holes,
      we have placed initially a classical accretion disk within the first 20 last stable radii,
      sandwiched by a hot and tenuous corona, and threaded by a large scale
     magnetic field. The solution procedure run as follows:
\ben
\item  The HD-equations are solved using the IOS-approach as depicted in Stage I
       of Fig. 3. The calculations were run to cover the viscous time scale.
\item  Using the obtained results from the previous stage as starting
       conditions, Stage II of the global
       solution procedure is now employed to run the calculations for an additional
       viscous time scale. Here,
       the HD and the MHD equations are solved in a blockwise manner as described in Fig. 3.
       Stage III was not employed, as {\Alfven}-waves propagation
       enhances the time-dependency of the flow even more.
\item       The final flow-configuration apparently governed by
       inflow and outflow plasmas. In general, outflows are gravitationally
       unbound, and therefore the corresponding Bernoulli number should be positive,
       whereas negative numbers correspond to gravitationally bound flows that
       should end their motion inside the black hole. Fig. 6 shows
       the 2D distribution of the Bernoulli number which obviously show the locations
       of the gravitationally-bound and unbound flows.
\een
\subsection{The spectral energy distribution of the in- and outflow
around the supermassive
       black hole of the giant elliptical galaxy M87}
       The results obtained in the previous subsection are used to construct the
       spectral energy distribution. Therefore, the last stage of the hierarchical scenario
       is now employed in combination with Stage II. Here, the solver of Stage II is
       activated once every several dozens iterations of the RT-solver.\\
       In Fig. 9 we display the results of several calculations under various conditions.
       The results displayed in Fig. 7 and 8 are preliminary, as the distributions of the
       density and temperature used here are artificial, but aimed at testing the convergence
       of the  RT-solver.
\section{The combined solution procedure: The hierarchical scenario}

In the following we describe the main steps of a possible
algorithmic procedure for solving the combined set of  MHD and the
RT equations (see Stages III and IV of Fig. 3): \ben
\item Compute the $RHS_\mm{i}$ and the Jacobian $A_\mm{i} = \partial Lq_\mm{i}/\D
q_\mm{i}$
       of each physical variable $q_\mm{i} (= \rho, m, n, ...)$, where
       $Lq_\mm{i}$ is the equation describing the evolution of variable $ q_\mm{i}.$
\item For each equation $Lq_\mm{i},$  compute the
      coefficient matrices $B_\mm{i} =  \partial Lq_\mm{i}/\partial q_\mm{j},$ for which
      $i\neq j$. This procedure applies for advection and diffusive operators only, though
      not for the source terms.
\item  Compute the coefficient matrices corresponding to the source terms only, i.e.,
       $H_\mm{i} = \partial Lq_\mm{i}/\partial q_\mm{j},$ for $i=1, N$ and
       $j=1,N$, and $i\neq j.$
\een
    The separation of the above-mentioned procedures is essential for enhancing the
    global efficiency of the hierarchical method. Specifically, the computation
    of each of the $B_i$ and $H_i$ is optional, depending on the problem in hand.
    For example, to solve the  system of equations corresponding to
    the hydrodynamical and isothermal flow in 1D efficiently, the numerical algorithm
    should be capable of calling the relevant routines only. Thus,  non-relevant
    routines can be switched off almost automatically, depending on the problem in hand.
    In particular, enlarging (reducing) dimensions, incorporating additional (excluding)
    variable  should be algorithmically maintainable.

    Taking into account that most  astrophysical flows are  of multi-scale by nature,
    we think that the hierarchical solution strategy might be a promising approach.
    In the following, we describe briefly the basis of this  hierarchical scenario
    applied to set of radiative MHD and the RT equations.

\ben
\item The hierarchical approach, or equivalently the multi-stage solution procedure,
      is based primarily on designing the global solver in such a manner to achieve
       maximum flexibility. Specifically,
      the numerical algorithm should be capable of solving the equations sequentially,
      block-sequential and/or in a fully-coupled manner.
      Re-ordering and using different pre-conditioning
      should be maintainable without changing the core of the inverter.
\item  As far as  vortex-free compressible, viscous and time-dependent flows are concerned, the implicit operator
       splitting approach (IOS) has been verified to be efficient and robust. IOS is most
       appropriate  for astrophysical fluid simulations, when the sought solutions
       depend weakly on the initial conditions, but strongly on the boundary conditions.
       The IOS-method is based on solving the set of equations sequentially as described
       in Stage I of Fig. 3.
       The convergence rate of the IOS-method may depend considerably on the order in which
       the equations are solved, provided the number of global iterations is low.
\item  The coupling between the equations can be enhanced gradually. From the cluster of
       coefficients, we may construct the Jacobians   $A^\mm{HD}$ and $A^\mm{MHD}$, which correspond
       to the set of HD and  MHD equations (see Stage II/Fig. 3). Algorithmically, this
       procedure is basically a sort of re-ordering and re-organizing of the coefficients, and
       does not require an extensive programming. As in the previous step, the order in which
       the equations are solved may affect both its convergence rate and efficiency.
       Here, a special care should be given to assure that the inclusion of  coefficients
       corresponding to the source terms does not enlarge the band width of  $A^\mm{HD}$ and
       $A^\mm{MHD}$. Test calculations
       have shown that careful ordering of the HD-equations may reduce the computational
       costs devoted for matrix inversion by
       $75\%$ \cite{HUJ1}. Furthermore,  it has been verified
       that several equations can still be separated and solved sequentially. Namely, the Possion equation
       for modelling self-gravity as well as the angular momentum equation accept partial
       decoupling from the rest of equations,
       provided the flow is axi-symmetric.
\item  Using the solutions obtained in stage II as initial conditions, we may solve the whole
       set of HD and MHD equations as a single set of coupled equations. The resulting Jacobian is
       highly sparse, for which pre-conditioned Krylov sub-iterative methods  are highly appropriate.

\item  By iterating over Stage II and IV, we can be sure that the resulting solution
       is reasonably close to sought quasi-stationary or steady solutions for the radiative MHD
       and radiative transfer equations. This is a consequence of:
\ben
\item  The radiative intensity in the high density regions, where the   optical thickness is large,
       is isotropic and coincides with black-body emission. Therefore, the intensity obtained
       by solving the zero moment of the radiation field is sufficiently accurate in this regime.
\item  The radiative intensity obtained by solving the RT-equation in optically thin regions may
       differ considerably from that obtained using the gray approximation. However, radiation in
       such regions have negligible power and they may hardly affect the dynamics of the flow.

      Consequently,   the following solution method may be  proposed:
      \bit
      \item The numerical values of the variables obtained in Stage III are used as
            initial conditions for calculating the non-gray and time-dependent radiative
            intensity.
      \item The mean-value of the frequency-dependent intensity is computed and subsequently
             used as initial condition for the radiative MHD equations.
      \item To avoid extensive computational costs, it is suggested to
            solve for $I_\nu$ every 10, or 20 time-steps. However, since the
            radiative time-scale is extremely short compared to the hydrodynamical
            time scale, it is much more reasonable to solve for the time-independent
            intensity.
      \eit

\een
\een
\section{Summary}

In this paper we have presented the hierarchical scenario  for solving
the set of radiative MHD equations and the 5D axi-symmetric radiative transfer
 equation. \\
The main features of this scenario are as follows: \ben
\item The global efficiency can be enhanced, depending on the optimal architecture of
      the global solver.  Specifically, the algorithmic structure should be sufficiently
      flexible, so that scalar or set of equations in arbitrary dimensions, different accuracies
      and using the appropriate pre-conditionings can be solved with a reasonable efficiency.
\item  Robustness is monitored through employing a  variety of solution procedures.
      Depending on the particular features of the problem considered, several stages
      of implicitness may be used, depending on the number of coefficients used
      for constructing the coefficient matrix. In particular, starting with a purely
      explicit time-stepping scheme, the algorithm should be capable of modifying the scheme into
      a fully implicit method dynamically.
\item For implicit calculations, the hierarchical algorithm relies on using a variety
      of preconditioning for accelerating convergence. For example, for modelling weakly
      incompressible flows, it has been verified that   the "Approximate
      Factorization Method" as pre-conditioning yields a larger convergence rate
      than  the ``Alternating Directional Implicit" or the "Line Gauss-Seidel" methods.
      However, the latter preconditionings provide faster convergence if the flow is
      compressible and advection-dominated. Therefore, depending on the problem in hand,
      the algorithm should be capable of employing the appropriate preconditioning
      at least in an explicit-adaptive manner.
\item The hierarchical algorithm is capable of solving  the angle-averaged
      time-dependent radiation transfer equation, taking into account the
      Kompaneets operator for modelling up-scattering of soft photons by hot
      electrons in magnetized plasmas.\\
      We note, however, that the assumption of isotropic radiative intensity  may break down
      if the flow is relativistic and contains regions  of significantly different optical
      depths.  Therefore, in the near future we intend to modify the
      RT-solver to enable modelling the motions of ultra-relativistic plasmas
      in the vicinities
      Kerr and Schwarzschild black holes.
\item The algorithm includes a procedure that allows solving  the zero-moment MHD
      equations partially/loosely coupled with the radiation transfer equations.
      The latter coupling can be significantly enhanced through  parallelization
      on powerful machines. 
\een

Finally, we have shown that the hierarchical algorithm presented
here can be applied
       to study the mechanisms underlying the formation, launching and acceleration of jets
       in AGNs and quasars, though serious numerical and physical modifications are still
        required.



\begin{thebibliography}{[MMMM]}
\bibitem[BA91]{Balbus}  Balbus, S., Hawley, J.,  1991, ``A powerful local shear instability in weakly magnetized disks. I - Linear
                         analysis. II - Nonlinear evolution",ApJ, \textbf{376}, 214
\bibitem[BW78]{Beam}Beam, R.M., Warming, R.F., 1978, ``An implicit factorized scheme for compressible Navier-Stokes equations'',AIAA, \textbf{16}, 393
\bibitem[BR01]{Brandt}Brandt, A., 2001, ``Textbook Multi-grids",in Multigrid, ed.: Trottenberg, U., Oosterlee, C., Sch\"uller, A.,
         Acad. Press, London
\bibitem[DO98]{Dongarra}Dongarra, I.S., Duff, D.C., Sorensen, H.A., van der Vorst, 1998, 
         ``Num. Linear Alg. for High-Performance Computers'',  SIAM J. Scient. Comput., \textbf{20}, 94
\bibitem[FE72]{Felten} Felten, J.E., \& Rees, M.J., 1972, ``Transfer effects on lines and continuum in optically thick sources'', \aap, \textbf{21}, 139  
\bibitem[FL88]{Fletcher} Fletcher, C.A.J., 1988, 'Computational Techniques for Fluid Dynamics',
                     Vol, I and II, Springer-Verlag
\bibitem[FO00]{Font}Font, J. A. 2000, ``Numerical Hydrodynamics in General Relativity",Living Rev. Relativity, \textbf{3}, 2
\bibitem[FR00]{Fryxell}Fryxell, B., Olson, K., Ricker, P., et al., 2000, ``FLASH: An Adaptive Mesh Hydrodynamics Code for Modeling
                        Astrophysical Thermonuclear Flashes", \apjs, \textbf{131}, 273
\bibitem[GA03]{Gammie}Gammie, C.F.,  McKinney, J.C., T${\rm\acute{o}}$th, G., 2003,  ``HARM: A Numerical Scheme for General Relativistic
                         Magnetohydrodynamics", \apj, \textbf{589}, 444
\bibitem[HI90]{Hirsch} Hirsch, C., 1990, 'Num. Computation of Internal and External Flows',
                      Vol, I, and II, John Wiley \& Sons, New York
\bibitem[HUJ0]{HUJ0}  Hujeirat, A., Papaloizou, J.C.P., 1998, ``Shock formation in accretion columns - a 2D radiative MHD approach'',\aap, \textbf{340}, 593
\bibitem[HUJ1]{HUJ1}  Hujeirat, A., Rannacher,R., 2001, ``On the efficiency and robustness of implicit methods
                                    in computational astrophysics'', NewAR, \textbf{45}, 425
\bibitem[HUJ2]{HUJ2} Hujeirat, A., Camenzind, M., Livio, M., 2002,
            ``Ion-dominated plasma and the origin of jets in quasars'', \aap, \textbf{394}, L9
\bibitem[HUJ3]{HUJ3}  Hujeirat, A., Camenzind, M., Burkert, A., 2002b,
        ``Comptonization and synchrotron emission in 2D accretion flows. I. A new numerical solver
                for the  Kompaneets equation'', \aap, \textbf{386}, 757
\bibitem[HUJ5]{HUJ5} Hujeirat, A., Livio, M., Camenzind, M., Burkert, A., 2003,
      ``A model for the jet-disk connection in BH accreting systems'', \aap, \textbf{408}, 415
\bibitem[HUJ6]{HUJ6} Hujeirat, A., Blandford, R.D., 2004,
  ``A model for electromagnetic extraction of rotational energy and formation
     of accretion-powered jets in radio galaxies'', \aap, \textbf{416}, 423
\bibitem[HUJ8]{HUJ8} Hujeirat, A., 2004, ``A method for enhancing the stability and robustness of 
     explicit schemes in CFD'', in press, New Astronomy Reviews.
\bibitem[KA76]{Katz}  Katz, J.A., 1976, ``Nonrelativistic Compton scattering and models of quasars'', ApJ, \textbf{206}, 910
\bibitem[IL72]{Iilarinov}  Iilarinov, A.F., \& Sunyaev, R.A., 1972, ``Compton scattering by thermal electrons in X-ray sources ''Soviet Astr. -AJ, \textbf{16}, 45
\bibitem[KL89]{Kley} Kley, W., 1989, ``Radiation hydrodynamics of the boundary layer in accretion disks. I -
                          Numerical methods", \aap, \textbf{208}, 98
\bibitem[KO99]{Koide1}Koide, S., Shibata, K., \& Kudoh, T. 1999, ``Relativistic Jet Formation from Black Hole Magnetized Accretion Disks:
                         Method, Tests, and Applications of a General
                         Relativistic Magnetohydrodynamic Numerical Code", \apj, \textbf{522}, 727
\bibitem[KO02]{Koide2}Koide, S., Shibata, K., Kudoh, T., \& Meier, D. L. 2002, ``Extraction of Black Hole Rotational Energy by a Magnetic Field and the
                         Formation of Relativistic Jets", Science, \textbf{195}, 1688
\bibitem[KO99]{Komissarov}Komissarov, S. S. 1999, ``A Godunov-type scheme for relativistic magnetohydrodynamics", \mnras, \textbf{303}, 343
\bibitem[LE81]{Levermore} Levermore, C.D., \& Pomraning, G.C.,  1981, ``A flux-limited diffusion theory'', ApJ, \textbf{248}, 321
\bibitem[MA96]{Mahadevan} Mahadevan, R., \& Narayan, R., Yi, I., 1996, `` Harmony of electrons: Cyclotron and Synchrotron emission by thermal electrons in magnetic fields'', ApJ, \textbf{465}, 327
\bibitem[MA85]{MacCormack} MacCormack, R.W., 1985, ``Current status of numerical solutions of Navier-Stokes equations'', AIAA, Paper 81-0110
\bibitem[MA99]{Mart}Mart${\rm\acute{i}}$, J.M., M\"uller, E., 1999, ``Numerical hydrodynamics in special relativity", Living Rev. Relativity, \textbf{2}, 3
\bibitem[ME01]{Meier1}Meier, D.L., Koide, S., \& Uchida, Y. 2001, ``Magnetohydrodynamic Production of Relativistic Jets", Science, \textbf{291}, 84
\bibitem[ME03]{Meier2}Meier, D., 2003, ``The theory and simulation of relativistic jet formation: towards a unified
                           model for micro- and macroquasars", NewAR, \textbf{47}, 667
\bibitem[MI86]{Mihalas} Mihalas, D., Mihalas, B.W., 1984, ``Foundations of radiation hydrodynamics'', Oxford University Press, NY, (MM)
\bibitem[OU97]{Ouyed} Ouyed, R., Pudritz, R., 1997, ``Numerical simulation of astrophysical jets from Keplerian disks. II. episodic outflows'', ApJ, \textbf{484}, 794 
\bibitem[PA80]{Payne} Payne, D.G., 1980, ``Time-dependent Comptonization - X-ray reverberations'', ApJ, \textbf{237}, 951
\bibitem[RY79]{Rybiki} Rybiki, G.B., \& Lightman, A.P., 1979, Radiation processes, Wiley-Interscience Publication
\bibitem[SA00]{Saad}Saad, Y., van der Vorst, 2000, ``Iterative solution of linear systems in the 20-th century'', J. of Comp. and Appl. Math., \textbf{123}, 1
\bibitem[SH76]{Shapiro} Shapiro, S.L., Lightman A.P., \& Eardley, D.M., ``A two-temperature accretion disk model for Cygnus X-1 structure and spectrum'',1976, ApJ, \textbf{204}, 187
\bibitem[ST92]{Stone}Stone, J.M., \& Norman, M., 1992, ``ZEUS-2D: A radiation magnetohydrodynamics code for astrophysical flows
                        in two space dimensions. I - The hydrodynamic algorithms and tests.", \apjs, \textbf{80}, 791
\bibitem[TO98]{Toth}T${\rm\acute{o}}$th, Keppens, R., Botchev, M.A., 1998, ``Implicit and semi-implicit schemes in the Versatile Advection Code:
                           numerical tests", \aap, \textbf{332}, 1159
\bibitem[TR01]{Trottenberg}Trottenberg, U., 2001, in Multigrid, ed.: Trottenberg, U., Oosterlee, C., Sch\"uller, A.,
         Acad. Press, London 
\bibitem[UC99]{Uchida}Uchida, Y., Nakamura, M., Hirose, S., Uemura, S., ``Magnetodynamic formation of jets in accretion process 
                             of magnetized mass onto the central gravitator'', Ap\&SS, \textbf{264}, 195
\bibitem[VI03]{Villiers}De Villiers, J.-P., \& Hawley, J.F., 2003, ``A Numerical Method for General Relativistic Magnetohydrodynamics", \apj, \textbf{589}, 458
\bibitem[ZI98]{Ziegler}Ziegler, U., 1998, ``NIRVANA+: An adaptive mesh refinement code for gas dynamics and MHD", Comp. Phys. Comm., \textbf{109}, 142
\end{thebibliography}
\end{document}